\pdfoutput=1
\documentclass[12pt]{article}
\usepackage{graphicx}
\usepackage{amssymb,amsmath,amsfonts,palatino,amsthm}
\usepackage{amssymb,mathtools}
\usepackage[sans]{dsfont}
\newcommand{\beq}{\begin{equation}}
\newcommand{\eeq}{\end{equation}}

\newcommand{\f}{\begin{equation}}
\newcommand{\ff}{\end{equation}}

\newtheorem{theorem}{Theorem}
\newtheorem{corollary}{Corollary}

\newtheorem{lemma}{Lemma}
\begin{document}

%%%%%%%%%%%%%%%%%%%%%%%%%%%%%%%%%%%%%%%%%%%%%%%%
\title{Conserved Quantities for Interacting Four Valent Braids in Quantum Gravity}
\author{
Jonathan Hackett\thanks{Email address:
jhackett@perimeterinstitute.ca} and Yidun Wan\thanks{Email address:
ywan@perimeterinstitute.ca}
\\
\\
\\
Perimeter Institute for Theoretical Physics,\\
31 Caroline st. N., Waterloo, Ontario N2L 2Y5, Canada, and \\
Department of Physics and Astronomy, University of Waterloo,\\
Waterloo, Ontario N2J 2W9, Canada\\}
\date{February 15, 2008}
\maketitle
\vfill
\begin{abstract}
We derive conservation laws from interactions of
actively-interacting braid-like excitations of embedded framed spin
networks in Quantum Gravity. Additionally we demonstrate that
actively-interacting braid-like excitations interact in such a way
that the product of interactions involving two actively-interacting
braid-like excitations produces a resulting actively-interacting
form.
\end{abstract}
\vfill
\newpage
\tableofcontents
\newpage

\section{Introduction}
There has been a large amount of research effort towards a quantum
theory of gravity with matter as topological invariants since a
ribbonized preon model\cite{Bilson-Thompson2005} was coded into
local braided ribbon excitations\cite{Bilson-Thompson2006}. These
excitations are edges of framed three-valent spin networks present
in models related to Loop Quantum Gravity with non-zero cosmological
constant\cite{Major1995}. The topological invariants of ribbon
braids are able to detect chirality and code chiral conservation
laws.

The results of \cite{Bilson-Thompson2006} have a serious
limitation, namely that the conservation laws which preserve the
excitations are exact. In other words, the braid excitations of
\cite{Bilson-Thompson2006} behave like solitons in integrable
systems: they propagate through each other under local dynamical
moves, leaving no possibility for interaction, e.g. creation and
annihilation, to occur\cite{Hackett2007}. Consequently, interpreting
braid excitations found in \cite{Bilson-Thompson2006} as particles -
in particular the Standard Model particles - is not going to work out
unless interactions are successfully introduced to that model.

Although there has been a search for a modification of the dynamics
studied in \cite{Bilson-Thompson2006} which would allow both
propagation and interactions of the local braid excitations, a new
model has also been put forward, which solves the problem of
interaction and opens new interesting directions worth of
investigation\cite{Wan2007,LeeWan2007}.

The new model extends the graphs from three-valent to four-valent
spin networks, embedded in a topological three-manifold, which also
gives rise to local braid excitations, and bases the dynamics on the
dual Pachner moves naturally associated with four-valent graphs. The
two main reasons of this extension are: that four-valent graphs and
the corresponding dual Pachner moves naturally occur in spin foam
models\cite{spin-foam}, and that vertices of four-valent spin
networks have true correspondence to three-dimensional space.
Nevertheless, a Pachner move in the model is allowed on a subgraph
only when the interior of the dual topology of the subgraph is
homeomorphic to an open trivial ball in $\mathbb{R}^3$. This
condition ensures the stability of certain braid
excitations\cite{LeeWan2007}.

This is the third of a series of papers along the research on the
new model. Our previous work formulates a clear and useful graphic
calculus as the tool to study the new model. We summarize our
previous main results as follows. In \cite{Wan2007} we found that
3-strand braids, each of which consists of two adjacent nodes
sharing three edges, are of most interest and defined equivalence
moves which relate projections of diffeomorphic embeddings of the
same spin network. Furthermore, we classified 3-strand braids into
two primary categories: reducible braids and irreducible braids. The
former contains braids which are equivalent to braids with less
number of crossings under equivalence moves. \cite{LeeWan2007} shows
that some stable braids propagate, in the sense that under the local
dynamical moves they exchange places with their adjacent
substructures in the graph. This propagation is in most cases
chiral, in which a braid can only propagates along an edge in the
larger graph only to the left and its mirror image (which is a
distinct state) can only propagates to the right. The chirality of
this propagation is related to our classification of braids in
\cite{Wan2007}. Additionally, two neighboring braids can interact,
in the sense that they merge into a new braid under local
equivalence and dynamical moves; however, for this to happen either
or both of them must be in a small class of braids called
actively-interacting. Lastly we found that any actively-interacting
braid is equivalent to some trivial braid with twists.

Due to the key role the actively-interacting braids play in this
model, in this paper we study their interactions in more details in
an algebraic way. Despite the power of the previously developed
graphic calculus in \cite{Wan2007,LeeWan2007}, a symbolic algebra
would provide a concise tool, convenient for computation and more
importantly more able to demonstrate conserved quantities as
"quantum numbers". As the ultimate goal of this model is to
interpret certain braid excitations as matter degrees of freedom in
a physical way - particularly to see if the braid excitations can be
mapped to Standard Model particles - we need a symbolic algebra
which exhibits possible symmetries in a more transparent and lucid
way. Indeed, as a follow-up work of \cite{Bilson-Thompson2006}, in
\cite{LouNumber} similar algebraic approach is being taken for
three-valent spin networks, in which each ribbon braid is only
characterized by three integers: the so-called Louis numbers, named
after Louis Kauffman. However, in the 4-valent case a braid is
characterized by more integral quantities, as will be seen.

To these ends within this paper, we shall develop a concise
algebraic notation of braids which interact actively. Within this
notation, the algebraic equivalence moves are defined, and the
quantities conserved under these are identified. Finally, the
algebra of interactions of actively-interacting braids is developed
in detail. From this the following results are found:
\begin{enumerate}
\item Conserved quantities exist under interactions and we are able to show the form of these conservation laws.
\item The interaction of two actively-interacting braids results in an
actively-interacting braid.
\end{enumerate}

It is worth noting that the algebraic notation proposed in this
paper will be used in the parallel papers\cite{HeWan2008b,
HeWan2008a, Wan2008}. \cite{HeWan2008b} extends the algebraic
notation and conserved quantities of actively-interacting braids to
all stable braids. Based on this, \cite{HeWan2008a} discovers the
discrete transformations of braids, which are mapped to C, P, T, and
their combinations. Then \cite{Wan2008} proposes an effective theory
of our braid excitations in terms of Feynman diagrams, which implies
the analogy between actively-interacting braids and bosons. This
implication conversely motivate this paper as well.

\section{Notation}%
\begin{figure}[h]
\begin{center}
\includegraphics[
natheight=0.909800in, natwidth=2.764800in, height=0.9426in,
width=2.808in ]{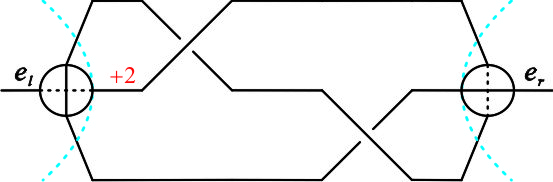}%
\caption{A typical 3-strand braid diagram formed by the three common
edges of two end-nodes. The region between the two dashed line
satisfies the definition of an ordinary braid. Edges $e_l$ and $e_r$
are called external edges. There is also a right handed twist of 2
units on the middle strand. In this figure the left end-node is in
an '$+$' state while the right
end-node is in an '$-$' state.}%
\label{braid}
\end{center}
\end{figure}

We will develop and use an algebraic notation, with, however,
keeping a graphic notation wherever necessary for an illustrative
purpose. We adopt the graphical notation we proposed in
\cite{Wan2007,LeeWan2007}. Fig. \ref{braid} shows an example of a
3-strand braid in this notation. More precisely, Fig. \ref{braid} is
a braid diagram which is a projection of the true 3-strand braid
embedded in a topological three manifold. Each spin network can be
embedded in various ways, some of which are diffeomorphic to each
other. The projection of a specific embedding of a braid is called a
braid diagram; many braid diagrams are equivalent and belong to the
same equivalence class, in the sense that they correspond to the
same braid and can be transformed into each other by equivalence
moves\cite{Wan2007}. Thus, a braid refers to the whole equivalence
class of its braid diagrams. Nevertheless, one can choose a braid
diagram of an equivalence class as the representative of the class,
we therefore will not distinguish a braid from a braid diagram in
the sequel unless an ambiguity arises. Besides, a braid always means
a 3-strand braid.

\begin{figure}
[h]
\begin{center}
\includegraphics[
natheight=1.909500in, natwidth=2.764800in, height=1.9476in,
width=2.808in
]%
{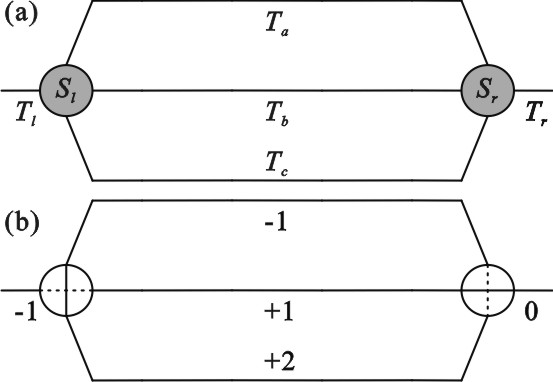}%
\caption{(a) is a trivial braid in general, in which the two nodes are %
filled in grey, indicating arbitrary states. (b) is a concrete example of (a).}%
\label{trivialBraid}%
\end{center}
\end{figure}

Among all the stable braids, a small class is called
\textbf{actively-interacting}, in the sense that they interact onto
any other braid as long as the interaction condition is met. These
braids are indeed "active" because they also propagate both to its
left and right\cite{LeeWan2007}, so they are able to find other
braids to interact. The interactions of passive braids, i.e. braids
which are not actively-interacting, are mediated by
actively-interacting braids, which implies the possibility that
actively-interacting braids behave bosonic\cite{Wan2008}. According
to \cite{LeeWan2007,Wan2007}, all actively interacting braids happen
to be both \textbf{completely left-} and \textbf{right-reducible},
i.e. such a braid can always be represented by a trivial braid
diagram with possibly twists on its three strands and two external
edges. This fact largely reduces the complexity of
actively-interacting braids because a trivial braid is fully
characterized by its twists and end-nodes' states, no crossings to
be worried about. Fig. \ref{trivialBraid}(a) depicts a trivial braid
in general.

Therefore, in this paper we concentrate on trivial braid diagrams
that represent actively-interacting braids. To study these braids in
more details and in a more efficient way, we need a nice algebraic
notation. We found that it is handy to denote a generic trivial
braid shown in Fig. \ref{trivialBraid}(a) algebraically by the
notation in Eq. \ref{braidRep},
\begin{equation}
_{T_l}^{S_l}\hspace{-0.5 mm}[ T_a,T_b,T_c]
_{T_r}^{S_r},\label{braidRep}
\end{equation}
where $S_l,S_r$, being '$+$ or '$-$', are respectively the states of
the left and right end-nodes of a braid, $T_l,T_r$, called the
\textbf{left} and \textbf{right external twists}, are respectively
the twists on the left and right external edges of a braid, and the
triple $[T_a, T_b, T_c]$ records the twists on the three strands
respectively in the order shown in Fig. \ref{trivialBraid}(a), which
are thus named the \textbf{internal twists}. The subscript $a$ of an
internal twist $T_a$ is abstract and has no meaning before its
position in the triple is fixed. So $[T_a, T_b, T_c]=[T_d, T_e,
T_f]$ means respectively $T_a=T_d,T_b=T_e$, and $T_c=T_f$. In the
rest of the paper, we will also consider the addition of two triple
of twists, i.e. $[T_a, T_b, T_c]+[T_d, T_e, T_f]=[T_a+T_d, T_b+T_e,
T_c+T_f]$.

All twists are valued in $\mathbb{Z}$ in units of
$\pi/3$\cite{Wan2007}. For example, the braid in Fig.
\ref{trivialBraid}(b) is denoted in the algebraic notation by
$_{-1}^{\ +}\hspace{-0.5 mm}[ -1,1,+2] _{0}^{-}$. Note that for a
trivial braid diagram that interact actively, the set
$\{T_l,S_l,T_a,T_b,T_c,S_r,T_r\}$ characterizing it is not
completely arbitrary but rather has the following constraints.
\begin{enumerate}
\item $S_l\equiv S_r$. If a braid is actively-interacting, each of
its trivial braid diagrams must have both end-nodes in the same
state (so the braid in Fig. \ref{trivialBraid}(b) does not actively
interact).
\item the triple $[T_a,T_b,T_c]$ is not arbitrary; however, the general
pattern of them, ensuring active interaction, has not yet been
found. Nevertheless, the algebra formulated in this paper may turn
out to be helpful to resolve this problem.
\end{enumerate}

\section{Algebra of equivalence moves: symmetries and relations}%
In the first of this series of papers\cite{Wan2007}, we discovered
equivalence moves on embedded 4-valent spin networks by which braid
diagrams are classified. Any braid diagram in an equivalence class
can be chosen to be the representative; however, for the purpose of
studying braid propagation and interaction, we found that each
equivalence class of braid diagrams has a unique representative
which has two twist-free external edges. If the unique
representative of a braid diagram interacts actively, so does the
whole class. Nevertheless, as mentioned above, in this paper we are
mainly interested in actively interacting braids which have trivial
braid diagrams; hence, it is now more convenient to represent an
equivalence class of braid diagrams by trivial braid diagrams in the
class which are certainly not unique. In other words, if a trivial
braid in our algebraic notation corresponds to a particle, there are
many other equivalent ones corresponding to the same particle. This
fundamental degeneracy of a particle is a direct result of
diffeomorphism invariance because equivalence moves do not change
the diffeomorphism of an embedding. Therefore, it is necessary to
find out for any equivalence class of braid diagrams, in an
algebraic way adapted to our new algebraic notation, how a trivial
representative is related to another in the same class, in
particular how the set of quantities characterizing the
representative changes, and more importantly what the conserved
quantities are.

\begin{figure}
[h]
\begin{center}
\includegraphics[
natheight=1.909500in, natwidth=2.809800in, height=1.9476in,
width=2.853in
]%
{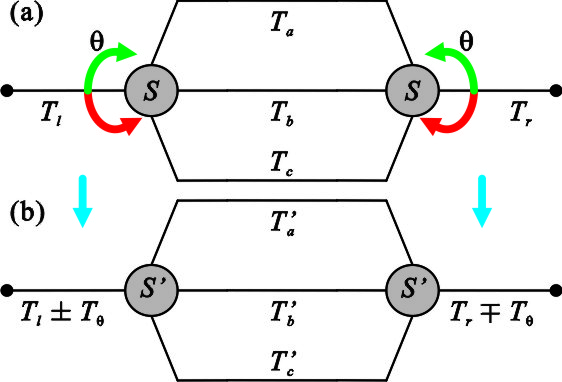}%
\caption{(a) is a actively interacting trivial braid in general; (b)
is obtained from (a) by same amount of rotations respectively on
both end-nodes of (a) in the same direction (either green or red)
shown in the figure. $T_\theta\in\mathbb{Z}$ is the twist induced by
the rotation $\theta$. $[T'_a,T'_b,T'_c]=[P^S_\theta(T_a,T_b,T_c)]$,
where $P^{S}_\theta$ is the permutation depending on both $S$ and $\theta$.}%
\label{rotation}%
\end{center}
\end{figure}
The tools we can utilize have already been introduced in
\cite{Wan2007} in detail, namely the rotations. That is, we can use,
e.g. $\pi/3$ rotations, to take a trivial braid to another.
Concentrating on trivial braids only, however, we do not expect any
crossing to appear because a single rotation on an end-node of a
braid diagram creates/annihilates crossings, so we must apply
rotations on both end-nodes of a trivial braid simultaneously, such
that no crossing arises. The idea is illustrated in Fig.
\ref{rotation}. A remark is that the "same direction" we mentioned
in the caption of Fig. \ref{rotation} is with respect to the surface
the braid is projected on, i.e. this piece of paper; nonetheless, in
our definition of rotation with respect to the rotation axis, i.e.
the external edges, the rotation $\theta$ of the left end-node and
the one of the right end-node are of opposite handedness. This has
two immediate consequences: 1)in Fig. \ref{rotation}(b), while the
resulted twist on the left external edge is $T_l\pm T_\theta$, the
one on the right is $T_r\mp T_\theta$, a sign difference appears;
and 2)this process creates no crossing and interestingly no extra
twists on the three strands of the braid but rather results in the
twists $[T'_a,T'_b,T'_c]$ in Fig. \ref{rotation}(b) as a permutation
$[P^S_\theta(T_a, T_b, T_c)]$ of the three twists in Fig.
\ref{rotation}(a). Note that $P^S_\theta$ depends also on $S$, the
state of the end-nodes before being rotated, which is a special
property of rotations in 4-valent case. We name a whole procedure of
this type on a braid a \textbf{simultaneous rotation}.

We can describe the action of a simultaneous rotation on a trivial
braid algebraically. We first define an operator for a simultaneous
rotation, which acts simultaneously on both end-nodes of braid, by
$R_{n,-n}$, where $n$ is the amount of rotation valued in units of
$\pi/3$, i.e. $n\in\mathbb{Z}$, and the signs, which must be
opposite in the first and the second subscripts, indicate
respectively the handedness of rotations on left and right
end-nodes. The second index of $R$ seems a bit redundant; however,
we keep it to make the rotation handedness explicit. Moreover, a
state $S$ is simply a sign, $+$ or $-$, so we let $-(+)=-$ and
$-(-)=+$, i.e. $-S=\bar{S}$ both denoting the opposite of $S$.
Hence, in general we have
\begin{equation}
R_{n,-n}(_{T_l}^{S}\hspace{-0.5 mm}[ T_a,T_b,T_c] _{T_r}^{S})=_{\ \
T_l+ n}^{(-)^nS}\hspace{-1.5 mm}[ P^S_n(T_a,T_b,T_c)]
_{T_r- n}^{(-)^nS},%
\label{rotAlgebraic}
\end{equation}
where $(-)^n$ enters the equation because a $\pi/3$ rotation changes
the end-node state once. When $n$ is even, $R_{n,-n}$ does not
change the end-node state of a braid, it is then christened an
\textbf{even simultaneous rotation}, otherwise it is named an
\textbf{odd simultaneous rotation}. The RHS of Eq.
\ref{rotAlgebraic} clearly presents the three effect of a
simultaneous rotation $R_{n,-n}$ has three effects on a trivial
braid: the change of end-node state depending on $n$, the change of
external twists depending on $n$, and the induced permutation
$P^S_n$ on the triple of internal twists, which is determined by the
end-node state $S$ before the action of $R$, $n$, and the handedness
of the rotation on the left end-node.

In view of that all possible rotations on a node of an embedded
4-valent spin network are generated by $\pi/3$
rotations\cite{Wan2007}, which correspond to $n=\pm 1$, $R$
satisfies
\begin{equation}
R_{n,-n}=R^{(n)}_{1,-1}, \label{powerRel}
\end{equation}
where $(n)$ means $n$-th power, e.g. $R^{(2)}_{1,-1}=R_{2,-2}$ and
$R^{(-2)}_{1,-1}=R_{-2,2}$. However, Eq. \ref{powerRel} is actually
formal and is not as simple as it appears to be. The reason lies in
the fact that a simultaneous rotation induces a permutation
depending on the end-node state before taking the rotation, and that
a $\pi/3$ inverses the end-node state. More precisely, for example,
a rotation $R_{2,-2}$ is equal to two $R_{1,-1}$ in a row; the first
$R_{1,-1}$ induces a permutation, flipping the end-node state from,
say $S$, to $\bar{S}$, so the permutation induced by the second
$R_{1,-1}$ now actually respects $\bar{S}$ rather than $S$. This
reads, mathematically,
$$R_{2,-2}(\left._{T_l}^{\ S}\hspace{-0.5 mm}[ T_a,T_b,T_c]
_{T_r}^{S}\right.)=R_{1,-1}(\left._{T_l+1}^{\hspace{5mm}
\bar{S}}\hspace{-0.5 mm}[ P^S_1(T_a,T_b,T_c)]
_{T_r-1}^{\bar{S}}\right.)=\left._{T_l+2}^{\hspace{5mm}
S}\hspace{-0.5 mm}[ P^{\bar{S}}_1P^S_1(T_a,T_b,T_c)]
_{T_r-2}^S\right.$$%
In other words, the permutation induced by $R_{2,-2}$ is
$P^{\bar{S}}_1P^S_1$ but not $P^S_1P^S_1$, we'll call this
permutation $P^S_2$. This is readily generalized to higher
rotations. Clearly, an induced permutation is an element of the
permutation group $S_3=Sym([T_a,T_b,T_c])$. In terms of disjoint
cycles,
$$S_3=\{\mathds{1},(1\ 2),(2\ 3),(1\ 3),(1\ 2\ 3),(1\ 3\ 2)\},$$ where for
example, $(1\ 2)$ means exchanging elements 1 an 2 in the triple
$[T_a,T_b,T_c]$ and leaving the third one fixed, and $(1\ 2\ 3)$
reads cyclicly moving the first element in the triple to the second,
the second to the third, and the third to the first. Moreover, for a
product of permutations, the order of its action on a triple is, in
our convention, from right to left.

\begin{figure}
[h]
\begin{center}
\includegraphics[
natheight=2.910100in, natwidth=2.809800in, height=2.9533in,
width=2.853in
]%
{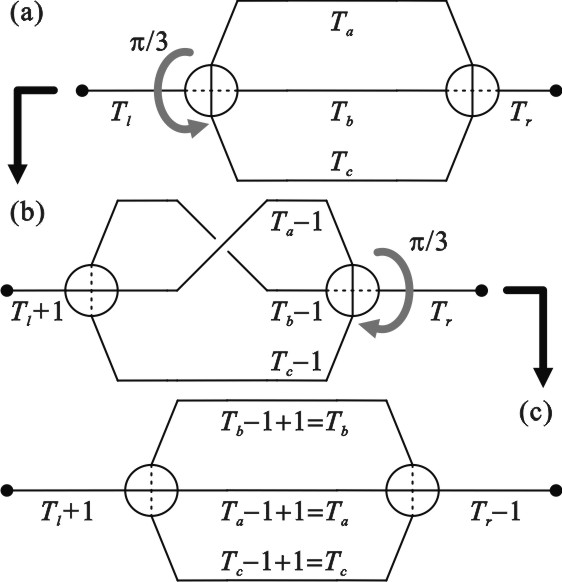}%
\caption{A simultaneous rotation of $\pi/3$ is split into two steps,
namely two $\pi/3$ rotations respectively on the two end-nodes of
the braid in (a). After a right-handed $\pi/3$ rotation on the left
end-node, (a) gives rises to (b), in which all twists but $T_r$ are
modified from those in (a) accordingly, and a crossing is created.
However, by another left-handed $\pi/3$ rotation on the right
end-node of (b), a trivial braid appears again in (c). One sees that
the induced permutation by this simultaneous rotation is
$P^+_{+1}=(1\ 2)$.}%
\label{rotPi3}%
\end{center}
\end{figure}

It is then demanded but sufficient to study a simultaneous rotation
of $\pi/3$ on both end-nodes of a trivial braid to fully understand
how a general simultaneous rotation works and to obtain our desired
algebraic relations in general. This is done in Fig. \ref{rotPi3}.

To obtain the exact algebraic form of this simultaneous rotation of
$\pi/3$, we can split the rotation into two rotations respectively
on the two end-nodes of the braid in Fig. \ref{rotPi3}. Our choice
is to first rotate left and then the right end-nodes of the trivial
braid in (a); it is obvious that the order of this splitting does
not matter, which guarantees that a simultaneous rotation is
well-defined. This rotation gives an equivalence relation between
the trivial braid in Fig. \ref{rotPi3}(a) and the one in (c), whose
algebraic form is now clear:
$$R_{+1,-1}(_{T_l}^{+}\hspace{-0.5 mm}[ T_a,T_b,T_c]
_{T_r}^{+})=\left._{T_l+1}^{\ \ \ \ -}\hspace{-0.5 mm}[ T_b,T_a,T_c]
_{T_r-1}^{-}\right..$$ The permutation implied in this relation is
$[P^+_{+1}(T_a,T_b,T_c)]=[T_b,T_a,T_c]$, i.e. $P^+_{+1}=(1\ 2)$.
Since each node has only two states and each rotation has only two
directions, it is not hard to enumerate all possible simultaneous
$\pi/3$ rotations on all possible actively interacting trivial
braids, in a way similar to what we do in Fig. \ref{rotPi3}. For
future convenience, we list all such relations in the following
table.
\begin{table}[h]
\begin{center}
\begin{tabular}
[c]{lll}%
$R_{+1,-1}(_{T_l}^{+}\hspace{-0.5 mm}[ T_a,T_b,T_c]
_{T_r}^{+})=\left._{T_l+1}^{\ \ \ \ -}\hspace{-0.5 mm}[
T_b,T_a,T_c] _{T_r-1}^{-}\right.$ &  & $P^+_{+1}=(1\ 2)$\smallskip\\%

$R_{-1,+1}(_{T_l}^{+}\hspace{-0.5 mm}[ T_a,T_b,T_c]
_{T_r}^{+})=\left._{T_l-1}^{\ \ \ \ -}\hspace{-0.5 mm}[
T_a,T_c,T_b] _{T_r+1}^{-}\right.$ &  & $P^+_{-1}=(2\ 3)$\smallskip\\%

$R_{+1,-1}(_{T_l}^{-}\hspace{-0.5 mm}[ T_a,T_b,T_c]
_{T_r}^{-})=\left._{T_l+1}^{\ \ \ \ +}\hspace{-0.5 mm}[
T_a,T_c,T_b] _{T_r-1}^{+}\right.$ &  & $P^-_{+1}=(2\ 3)$\smallskip\\%

$R_{-1,+1}(_{T_l}^{-}\hspace{-0.5 mm}[ T_a,T_b,T_c]
_{T_r}^{-})=\left._{T_l-1}^{\ \ \ \ +}\hspace{-0.5 mm}[
T_b,T_a,T_c] _{T_r+1}^{+}\right.$ &  & $P^-_{-1}=(1\ 2)$\\%
\end{tabular}%
\end{center}%
\caption{All possible simultaneous $\pi/3$ rotations on trivial braids which interact actively.}%
\label{rotPi3Algebra}
\end{table}

From Table \ref{rotPi3Algebra}, one can see that the following
quantities are invariant under these simultaneous
rotations:

\begin{enumerate}
\item $T_l+T_r$;
\item $T_a$, $T_b$, and $T_c$ are conserved individually modulo
permutation;
\item $S^2$ (which additionally means that the interacting nature of braids is preserved under the simultaneous rotations).
\end{enumerate}

Furthermore, because of Eq. \ref{powerRel}, this result applies to
any simultaneous rotation. There are also two derived conserved
quantities: specifically $T_a+T_b+T_c$ and
$T_{\text{total}}=T_l+T_a+T_b+T_c+T_r$. $T_{\text{total}}$ is
consistent to the overall conserved quantity found in
\cite{Wan2007}: the sum of all twists and crossings of a subgraph -
in particular a braid diagram - under any equivalent move. For a
trivial braid no crossing exists and so the overall conserved
quantity consists of twists only, this is $T_{\text{total}}$. All
these conserved quantities are invariants under diffeomorphic
embeddings.

Relations in Table \ref{rotPi3Algebra} are generating relations for
all possible simultaneous rotations, for example,
\begin{align*}
R_{+3,-3}(_{T_l}^{+}\hspace{-0.5 mm}[ T_a,T_b,T_c] _{T_r}^{+})
&=R_{+2,-2}(_{T_l+1}^{\ \ \ \ -}\hspace{-0.5 mm}[
T_b,T_a,T_c] _{T_r-1}^{-})\\
&=R_{+1,-1}(_{T_l+2}^{\ \ \ \ +}\hspace{-0.5 mm}[
T_b,T_c,T_a] _{T_r-2}^{+})\\
&=_{T_l+3}^{\ \ \ \ -}\hspace{-1.5 mm}[ T_c,T_b,T_a] _{T_r-3}^{-},
\end{align*}
where the $\pi$ rotation is realized by three consecutive $\pi/3$
rotations. It is easy to check by our graphic calculus that the
above calculation is indeed correct. Note that one must be careful
of performing an induced permutation at each step in such a
calculation as permutation depends on both the end-node state before
the rotation and the rotation handedness. In particular we will see
three important identities regarding the induced permutations
shortly. One can observe from Table \ref{rotPi3Algebra} that
\begin{equation}
\begin{aligned}
&P^S_{\pm 1}P^{\bar{S}}_{\mp
1}=\mathds{1}\\
&P^S_{\pm 1}=P^{\bar S}_{\mp 1},
\end{aligned}\label{permuRelproton}
\end{equation}
where $S$ is the end-node state and $\bar{S}$ is its opposite. This
can actually be generalized to the following lemma:
\begin{lemma}
\begin{align}
P^S_{2n}P^S_{-2n} &\equiv\mathds{1}\label{permuRelEven}\\
P^S_{2n+1}P^{\bar{S}}_{-(2n+1)} &\equiv\mathds{1}\label{permuRelOdd}\\
P^S_{n} &\equiv P^{\bar S}_{-n}, \label{permuRel}%
\end{align}
where $n\in\mathbb{Z}$.
\end{lemma}
\begin{proof}
 We shall begin by proving Eq. \ref{permuRelEven} (Eq.
\ref{permuRelOdd} follows similarly). First we have
$$P^S_{\pm 2}P^S_{\mp 2}=P^{\bar{S}}_{\pm 1}P^S_{\pm 1}P^{\bar{S}}_{\mp
1}P^S_{\mp 1}=P^{\bar{S}}_{\pm 1}\mathds{1}P^S_{\mp 1}=\mathds{1},$$
where the last two equalities hold by repeatedly applying Eq.
\ref{permuRelproton}, and $\bar{S}$ appears as a single
$\pi/3$ rotation flips $S$. Using this we obtain
\begin{align*}
P^S_{\pm 2n}P^S_{\mp 2n} &=(P^S_{\pm 2})^n(P^S_{\mp 2})^n\\
&=(P^S_{\pm 2})^{n-1}\underset{=\mathds{1}}{\underbrace{P^S_{\pm 2}P^S_{\mp 2}}}(P^S_{\mp 2})^{n-1}\\
&=(P^S_{\pm 2})^{n-1}(P^S_{\mp 2})^{n-1}\\
&=\mathds{1},
\end{align*}
where the last line results from repeatedly performing the same expansion as in
the second line. We then prove Eq. \ref{permuRel} by induction. It
is already true for $n=\pm 1$ (Eq. \ref{permuRelproton}). Assume it
holds for $-(k-1)\leq n\leq k-1$ for some $k\in\mathbb{N}$, then we
have for $|n|=k$
\begin{equation*}
P^S_{\pm k}=
\begin{cases} P^{\bar S}_{\pm 1}P^S_{\pm(k-1)}=P^S_{\mp 1}P^{\bar S}_{\mp(k-1)}=P^{\bar S}_{\mp k} & \text{if $k$ is even,}
\\
P^S_{\pm 1}P^S_{\pm(k-1)}=P^{\bar S}_{\mp 1}P^{\bar
S}_{\mp(k-1)}=P^{\bar S}_{\mp k} &\text{if $k$ is odd.}
\end{cases}
\end{equation*}
Thus Eq. \ref{permuRel} holds. The equations below are easy to
derive; they are listed here for possible future use.
\begin{equation}
\begin{aligned}
P^+_2=P^-_{-2} &=(1\ 3\ 2)\\
P^+_{-2}=P^-_2 &=(1\ 2\ 3)\\
P^{\pm}_{6n+3} &\equiv(1\ 3)\\
P^{\pm}_{6n} &\equiv\mathds{1},
\end{aligned}
\end{equation}
where $n\in\mathbb{Z}$.
\end{proof}

\section{Algebra of interactions: symmetries and relations}
Now that we consider trivial braids which interact actively, we may
ask a question: do two actively interacting braids always interact
to form another actively interacting braid? To answer this question,
we need to first find out if two adjacent actively interacting
braids, say $B$ and $B'$, always interact. One may feel this
question somehow bizarre at first glance because how it is possible
that two actively interacting braids do not interact.
\begin{figure}
[h]
\begin{center}
\includegraphics[
natheight=0.909800in, natwidth=2.945600in, height=0.9426in,
width=2.9888in
]%
{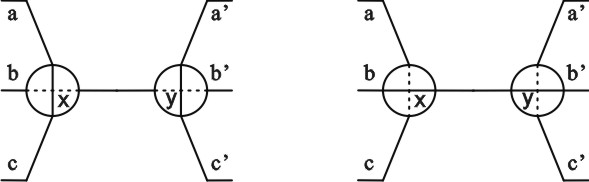}%
\caption{The only two possible configurations allowing a
$2\rightarrow3$
move.}%
\label{2to3}%
\end{center}
\end{figure}

Nevertheless, the reason to ask such a question is two-fold.
Firstly, the first step of performing an interaction of $B$ and $B'$
is doing a $2\rightarrow3$ move about the two adjacent end-nodes of
$B$ and $B'$ respectively; however, a $2\rightarrow3$ move is doable
only if the two adjacent nodes are of the same state and their
common edge is twist-free\cite{LeeWan2007}, as shown in Fig.
\ref{2to3}. Secondly, in our previous work, we chose to represent an
equivalence class of braid diagrams by its unique representative
which has twist-free external edges but in this paper we represent a
class by its trivial braid diagrams whose external edges are not
necessary twist-free. Therefore, when two trivial braids $B$ and
$B'$ meet each other, to see if they interact we should first check
if we can put their adjacent end-nodes in the same state and with a
twist-free common edge in between, which constructs the
\textbf{interaction condition}. In search of the answer to this
question, an algebra of interactions between actively-interacting
braids will be naturally developed.
\begin{figure}
[h]
\begin{center}
\includegraphics[
natheight=1.909500in, natwidth=2.809800in, height=1.9476in,
width=2.853in
]%
{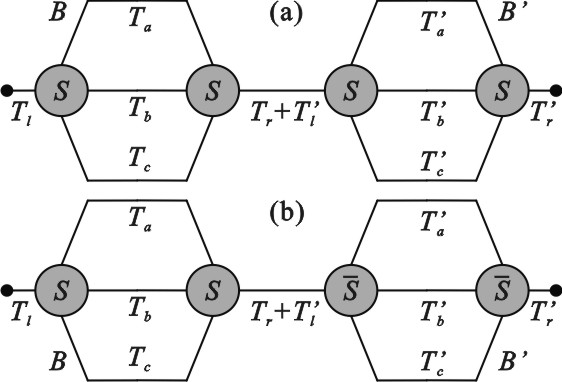}%
\caption{Two general cases in which two actively interacting braids
$B$ and $B^{\prime}$ meet each other: (a) $B$ and $B^{\prime}$ have
adjacent end-nodes in the same state; and (b) $B$ and $B^{\prime}$
have adjacent end-nodes in opposite states. $\bar{S}$ stands for the
opposite of $S$. The interaction condition in cases (a) and (b) are
respectively $T_r+T'_l=2n$ and $T_r+T'_l=2n+1$, $n\in\mathbb{Z}$.}
\label{connectedsum}%
\end{center}
\end{figure}

For two actively-interacting trivial braids $B$ and $B'$ with $B$ on
the left of $B'$, there are two general cases to study, which are
illustrated in Fig. \ref{connectedsum}. In Fig.
\ref{connectedsum}(a), the right end-node of $B$ and the left
end-node of $B'$ are already in the same state; hence, the task is
not only to get rid of the overall twist $T_r+T'_l$ on the common
edge but also to keep the adjacent end-nodes in the same state. We
know that only multiples of $2\pi/3$ rotations do not change the
state of the node being rotated, we thus demand that in this case
$T_r+T'_l=2n,n\in\mathbb{Z}$. For example, if $T_r+T'_l=2$, we can
perform a left-handed $2\pi/3$ rotation either on the right end-node
of $B$ or on the left end-node of $B'$, such that the two adjacent
end-nodes are still in state $S$ and their common edge is now
twist-free.

As to Fig. \ref{connectedsum}(b), the right end-node of $B$ and the
left end-node of $B'$ are in opposite states, so we should rotate
either of them to set them in the same state and annihilate the
twist $T_r+T'_l$ at the mean time. Based on the fact that only odd
multiples of $\pi/3$ rotations flip the state of the node under
rotation, we clearly must have in this case,
$T_r+T'_l=2n+1,n\in\mathbb{Z}$.

There is a good news. As long as the above interaction condition
holds, such that a $2\rightarrow3$ move can be done on the two
neighboring end-nodes of $B$ and $B'$, according to our definitions
of interaction and of actively-interacting braids, $B$ and $B'$ can
definitely interact to form another braid, written as $B+B'$, which
is also actively interacting. The reason is understood this way: the
interaction, left or right, of $B+B'$ onto an arbitrary braid
(interacting or not), say $B''$, can always be thought as a
composite process of two interactions, namely
$(B+B')+B''=B+(B'+B'')$, in which $B$ is the active braid in the
second step, or $B''+(B+B')=(B''+B)+B'$, in which $B'$ is the active
one in the second step. In other words, an interaction is
associative, with however, an active braid always playing the active
role.

Nevertheless, how does $B+B'$ look like as represented by a trivial
braid diagram? To perform the interaction, either $B$'s right
end-node or $B'$'s left end-node may be rotated before one can do a
$2\rightarrow3$ move, which disguises the form of $B+B'$ from being
guessed directly. Fortunately, this issue can be tackled as follows.

Let us study the case in Fig. \ref{connectedsum}(a). Supposing the
condition, $T_r+T'_l=2n$, holds, we have two subcases. The first is
the significantly easier to prove and so we shall address it first:
\begin{lemma}
Given two actively interacting braids $B=\left. _{T_l}^S\hspace{-0.5 mm}[ T_a,T_b,T_c]
_{T_r}^S\right.$ and $B'=\left. _{\ T_l'}^{\pm S}\hspace{-0.5 mm}[
T'_a,T'_b,T'_c] _{T'_r}^{\pm S}\right.$ satisfying the interaction conditions with $T_r+T'_l=0$, the interaction of $B$ and $B'$ gives $B^f=\left._{T_l}^S\hspace{-0.5 mm}[ (T_a,T_b,T_c) + (T'_a,T'_b,T'_c)]
_{T'_r}^S =\hspace{1 mm}_{T_l}^S\hspace{-0.5 mm}[
T_a+T'_a,T_b+T'_b,T_c+T'_c] _{T'_r}^S\right.$
\end{lemma}
\begin{proof}
As $T_r+T'_l=0$, no rotation is needed; hence, according to
\cite{LeeWan2007}, $B+B'$ forms a connected sum of $B$ and $B'$,
which reads, in our algebraic language,
\begin{equation}
\begin{aligned}
B+B'\overset{T_r+T'_l=0}{=}B \# B' &=\left._{T_l}^S\hspace{-0.5 mm}[
T_a,T_b,T_c] _{T_r}^S\right. \#\left. _{-T_r}^{\ \ \ S}\hspace{-0.5
mm}[ T'_a,T'_b,T'_c] _{T'_r}^S\right.\\
&=\left._{T_l}^S\hspace{-0.5 mm}[ (T_a,T_b,T_c) + (T'_a,T'_b,T'_c)]
_{T'_r}^S =\hspace{1 mm}_{T_l}^S\hspace{-0.5 mm}[
T_a+T'_a,T_b+T'_b,T_c+T'_c] _{T'_r}^S\right..
\end{aligned}\label{connectedsumAlg}
\end{equation}
\end{proof}
Before dealing with the other subcase, it is useful to
prove a Lemma.
\begin{lemma}
A simultaneous rotation commutes with a connected sum. In algebraic
words, this reads
\begin{equation}
R_{n,-n}(B\# B')=R_{n,-n}(B)\# R_{n,-n}(B'),\
n\in\mathbb{Z}.\label{theoCommAlg}
\end{equation}
\label{theoComm}
\end{lemma}
\begin{proof}
Let $B=\left._{T_l}^S\hspace{-0.5 mm}[ T_a,T_b,T_c] _{T_c}^S\right.$
and $B'=\left._{-T_c}^{\ \ \ S}\hspace{-0.5 mm}[ T'_a,T'_b,T'_c]
_{T'_r}^S\right.$. Note that $B$ and $B'$ have the same end-node
state, and that the right external twist of $B$, $T_c$, cancels the
left external twist of $B'$, $-T_c$, which is the most general
situation in which a connected sum is viable. Then, we have
\begin{align*}
R_{n,-n}(B)\# R_{n,-n}(B') &=R_{n,-n}(_{T_l}^S\hspace{-0.5 mm}[
T_a,T_b,T_c] _{T_c}^S)\# R_{n,-n}(_{-T_c}^{\ \ \ S}\hspace{-0.5
mm}[ T'_a,T'_b,T'_c] _{T'_r}^S)\\
&=\left._{\ T_l+n}^{(-)^nS}\hspace{-0.5 mm}[ P^S_{n}(T_a,T_b,T_c)]
_{T_c-n}^{(-)^nS}\right.\#\hspace{1 mm}
_{-T_c+n}^{\ (-)^nS}\hspace{-0.5 mm}[ P^S_{n}(T'_a,T'_b,T'_c)] _{T'_r-n}^{(-)^nS}\\
%\overset{T_c-n+(-T_cn)=0}{\Longrightarrow}
\xRightarrow{T_c-n+(-T_cn)=0}&=\left._{\ T_l+n}^{(-)^nS}\hspace{-0.5
mm}[ P^S_{n}(T_a+T'_a,T_b+T'_b,T_c+T'_c)] _{T'_r-n}^{(-)^nS}\right.\\
&=R_{n,-n}(_{T_l}^S\hspace{-0.5 mm}[
T_a+T'_a,T_b+T'_b,T_c+T'_c] _{T'_r}^S)\\
\overset{Eq. \ref{connectedsumAlg}}{\Longrightarrow}&=R_{n,-n}(B\#
B'),
\end{align*}
closing the proof.
\end{proof}
If $T_r+T'_l=2n\neq0$, we are in the second subcase. We can choose
to do a rotation of $-(T_r+T'_l)$ on either the right end-node of
$B$ or on the left end-node of $B'$ to annihilate the twist
$T_r+T'_l$ on the common edge. If we rotate the right end-node of
$B$ by $-(T_r+T'_l)$, we create crossings on $B$ and change $B$'s
twists, but we want to stay with the trivial braid diagrams. To get
around of this problem we simply need to perform a rotation of the
opposite handedness on the left end-node of $B$ as well, such that
$B$ is still in its trivial representation. In other words, we
perform a simultaneous rotation of $T_r+T'_l$ on $B$. Likewise, if
we rotate the left end-node of $B'$ to cancel the twist $T_r+T'_l$,
we should perform a simultaneous rotation of $-(T_r+T'_l)$ on $B'$.
In either choice, we arrive at a situation legal for a connected
sum. Nonetheless, each choice yields a result for $B+B'$, a natural
question is whether the two results "the same", or to be precise,
equivalent? The answer is "Yes", which is not only true for this
case but also valid for the case in Fig. \ref{connectedsum}(b).

We now prove this statement as a theorem. Here is the setup of the
theorem. For two actively-interacting trivial braids $B$ and $B'$,
with $B$ adjacent to $B'$ on the left, if the interaction condition
holds, one may need to perform a simultaneous rotation on either $B$
or $B'$ to remove the twist on their common edge for the interaction
to happen by a connected sum.
\begin{theorem}
The resulting interaction $B+B'$ between two actively interacting
braids does not depend on the choice of the braid being
simultaneously rotated.
\end{theorem}
\begin{proof}
Let $B=\left. _{T_l}^S\hspace{-0.5 mm}[ T_a,T_b,T_c]
_{T_r}^S\right.$ and $B'=\left. _{\ T_l'}^{\pm S}\hspace{-0.5 mm}[
T'_a,T'_b,T'_c] _{T'_r}^{\pm S}\right.$, which considers both cases
in Fig. \ref{connectedsum}: $T_r+T'_l=2n,n\in\mathbb{Z}$ for $+S$,
and $T_r+T'_l=2n+1,n\in\mathbb{Z}$ for $-S$. Then, $B+B'=\left.
_{T_l}^S\hspace{-0.5 mm}[ T_a,T_b,T_c] _{T_r}^S\right. +\left. _{\
T_l'}^{\pm S}\hspace{-0.5 mm}[ T'_a,T'_b,T'_c] _{T'_r}^{\pm
S}\right.$, and we have
\begin{align}
B+B' &=\left._{T_l}^S\hspace{-0.5 mm}[ T_a,T_b,T_c] _0^S\right.\#
R_{-(T_r+T'_l),T_r+T'_l}\left(_{T_r+T_l'}^{\ \ \ \ \pm
S}\hspace{-0.5
mm}[ T'_a,T'_b,T'_c] _{T'_r}^{\pm S}\right)\nonumber\\
&=\left._{T_l}^S\hspace{-0.5 mm}[ T_a,T_b,T_c] _0^S\right.\#\left.
_{\hspace{16 mm} 0}^{(-)^{T_r+T_l'}
(\pm S)}\hspace{-0.5 mm}[ P^{\pm S}_{-(T_r+T_l')}(T'_a,T'_b,T'_c)] _{T'_r+T_r+T_l'}^{(-)^{T_r+T_l'}(\pm S)}\right.\nonumber\\
\xRightarrow{(-)^{T_r+T_l'} (\pm S)=S}&=\left._{T_l}^S\hspace{-0.5
mm}[ T_a,T_b,T_c] _0^S\right.\#\left.
_{0}^{S}\hspace{-0.5 mm}[ P^{\pm S}_{-(T_r+T_l')}(T'_a,T'_b,T'_c)] _{T'_r+T_r+T_l'}^{S}\right.\nonumber\\
&=\left._{T_l}^S\hspace{-0.5 mm}[(T_a,T_b,T_c)+(P^{\pm
S}_{-(T_r+T_l')}(T'_a,T'_b,T'_c))]
_{T'_r+T_r+T_l'}^{S}\right.,\label{theoIntRchoice}
\end{align}
\begin{align}
B+B' &=R_{T_r+T'_l,-(T_r+T'_l)}\left(_{T_l}^S\hspace{-0.5 mm}[
T_a,T_b,T_c] _{T_r+T'_l}^S\right)\#\left. _{\ \ 0}^{\pm
S}\hspace{-0.5 mm}[ T'_a,T'_b,T'_c] _{T'_r}^{\pm
S}\right.\nonumber\\
&=\left._{\ T_l+T_r+T'_l}^{(-)^{T_r+T_l'}S}\hspace{-0.5 mm}[
P^S_{T_r+T_l'}(T_a,T_b,T_c)] _0^{(-)^{T_r+T_l'}S}\right.\#\left. _{\
\ 0}^{\pm S}\hspace{-0.5 mm}[ T'_a,T'_b,T'_c] _{T'_r}^{\pm S}\right.\nonumber\\
\xRightarrow{(-)^{T_r+T_l'} S=\pm S}
&=\left._{T_l+T_r+T'_l}^{\hspace{9mm} \pm S}\hspace{-0.5 mm}[
(P^S_{T_r+T_l'}(T_a,T_b,T_c))+(T'_a,T'_b,T'_c)] _{T'_r}^{\pm
S}\right.,\label{theoIntLchoice}
\end{align}
and
\begin{align}
&\left._{T_l}^S\hspace{-0.5 mm}[(T_a,T_b,T_c)+(P^{\pm
S}_{-(T_r+T_l')}(T'_a,T'_b,T'_c))]
_{T'_r+T_r+T_l'}^{S}\right. \nonumber\\
\cong\hspace{2mm} &\left._{T_l+T_r+T'_l}^{\hspace{9mm} \pm
S}\hspace{-0.5 mm}[ (P^S_{T_r+T_l'}(T_a,T_b,T_c))+(T'_a,T'_b,T'_c)]
_{T'_r}^{\pm S}\right., \label{theoIntLR}
\end{align}
where $\cong$ means "equivalent to". Note that $\pm S$ stands for the two cases respectively.%
\label{theoInt}

It is sufficient to prove Eq. \ref{theoIntLR}. In view of Lemma
\ref{theoComm}, we may apply a simultaneous rotation of $T_r+T'_l$
on the LHS of Eq. \ref{theoIntLR}, but perform the rotation
respectively on the two components of the connected sum and then
take the connected sum. That is,
\begin{align*}
& \left._{T_l}^S\hspace{-0.5 mm}[(T_a,T_b,T_c)+(P^{\pm
S}_{-(T_r+T_l')}(T'_a,T'_b,T'_c))]
_{T'_r+T_r+T_l'}^{S}\right.\\
\cong & R_{T_r+T'_l,-(T_r+T'_l)}\left( \left._{T_l}^S\hspace{-0.5
mm}[ T_a,T_b,T_c] _0^S\right.\#\left.
_{0}^{S}\hspace{-0.5 mm}[ P^{\pm S}_{-(T_r+T_l')}(T'_a,T'_b,T'_c)] _{T'_r+T_r+T_l'}^{S}\right. \right)\\
\overset{Eq. \ref{theoCommAlg}}{\Longrightarrow} =&
R_{T_r+T'_l,-(T_r+T'_l)}\left( \left._{T_l}^S\hspace{-0.5
mm}[ T_a,T_b,T_c] _0^S\right. \right)\\
& \#\hspace{1mm} R_{T_r+T'_l,-(T_r+T'_l)}\left( \left.
_{0}^{S}\hspace{-0.5 mm}[ P^{\pm S}_{-(T_r+T_l')}(T'_a,T'_b,T'_c)] _{T'_r+T_r+T_l'}^{S}\right. \right)\\
=& \left. _{\ T_l+T_r+T'_l}^{(-)^{T_r+T_l'}S}\hspace{-0.5 mm}[
P^S_{T_r+T_l'}(T_a,T_b,T_c)] _{-(T_r+T'_l)}^{(-)^{T_r+T_l'}S}\right.\\
& \#\left. _{\hspace{6.5 mm} T_r+T'_l}^{(-)^{T_r+T_l'} S}
\hspace{-0.5 mm}[ P^{(-)^{T_r+T_l'} S}_{T_r+T_l'}P^{\pm
S}_{-(T_r+T_l')}(T'_a,T'_b,T'_c)]
_{T_r'}^{(-)^{T_r+T_l'}S}\right.\\
=& \left. _{\ T_l+T_r+T'_l}^{(-)^{T_r+T_l'}S}\hspace{-0.5 mm}[
P^S_{T_r+T_l'}(T_a,T_b,T_c)] _{-(T_r+T'_l)}^{(-)^{T_r+T_l'}S}
\right.\#\left. _{T_r+T'_l}^{\ \ \ \ \pm
S}\hspace{-0.5 mm}[ T'_a,T'_b,T'_c] _{T_r'}^{\pm S}\right.\\
=& \left. _{T_l+T_r+T'_l}^{\hspace{9mm} \pm S}\hspace{-0.5 mm}[
P^S_{T_r+T_l'}(T_a,T_b,T_c)] _0^{\pm S} \right.\#\left.
_{\ \ 0}^{\pm S}\hspace{-0.5 mm}[ T'_a,T'_b,T'_c] _{T_r'}^{\pm S}\right.\\
=& \left. _{T_l+T_r+T'_l}^{\hspace{9mm} \pm S}\hspace{-0.5 mm}[
(P^S_{T_r+T_l'}(T_a,T_b,T_c))+(T'_a,T'_b,T'_c)] _{T'_r}^{\pm
S}\right.,
\end{align*}
which is the very Eq. \ref{theoIntLR}. The second to the last line
holds because the right subscript of the first term, $-(T_r+T'_l)$,
cancels the left subscript of the second term, $T_r+T'_l$, and
$(-)^{T_r+T_l'}= \pm S$ respectively in the two cases. The third
equality is valid for that
\begin{equation}
P^{(-)^{T_r+T_l'} S}_{T_r+T_l'}P^{\pm S}_{-(T_r+T_l')}=\left\{ %
\begin{array}
[c]{l}%
P^S_{T_r+T_l'}P^S_{-(T_r+T_l')}\overset{Eq. \ref{permuRelEven}}{=}\mathds{1},\ \ T_r+T_l'=2n \medskip\\
P^{-S}_{T_r+T_l'}P^{-S}_{-(T_r+T_l')}\overset{Eq.
\ref{permuRelEven}}{=} \mathds{1}, \ \ T_r+T_l'=2n+1
\end{array}
\right.\nonumber
\end{equation}
This completes the proof.
\end{proof}
\begin{figure}
[h]
\begin{center}
\includegraphics[
natheight=0.909800in, natwidth=2.760500in, height=0.9426in,
width=2.8029in
]%
{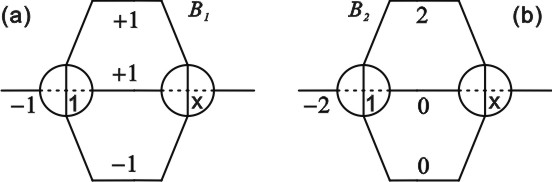}%
\caption{Two braids which interact actively.}%
\label{exampleInt}%
\end{center}
\end{figure}
Let us see an example. Fig. \ref{exampleInt} shows two
actively-interacting trivial braids\cite{LeeWan2007}, algebraically
they are $B_1=_{-1}^{\ +}\hspace{-1.5mm}[1,1,-1]_0^+$ and
$B_2=_{-2}^{\ \ +}\hspace{-1.5mm}[T+2,0,0]_0^+$. One sees that
\begin{align*}
B_1+B_2 &=_{-1}^{\ +}\hspace{-1.5mm}[1,1,-1]_0^+ + _{-2}^{\ \
+}\hspace{-1.5mm}[2,0,0]_0^+\\
&= _{-1}^{\ +}\hspace{-1.5mm}[1,1,-1]_0^+ \hspace{1mm}\#\hspace{1mm}
R_{2,-2}(_{-2}^{\ \
+}\hspace{-0.5mm}[2,0,0]_0^+)\\
&= _{-1}^{\ +}\hspace{-1.5mm}[1,1,-1]_0^+
\hspace{1mm}\#\hspace{1mm} _{\hspace{0.4mm}0}^+\hspace{-0.5mm}[P^+_2(2,0,0)]_{-2}^+\\
&= _{-1}^{\ +}\hspace{-1.5mm}[1,1,-1]_0^+ \hspace{1mm}\#\hspace{1mm}
_{\hspace{0.4mm}0}^+\hspace{-0.5mm}[0,0,2]_{-2}^+\\
&=_{-1}^{\ +}\hspace{-1.5mm}[1,1,1]_{-2}^+.
\end{align*}
On the other hand,
\begin{align*}
B_1+B_2 &=_{-1}^{\ +}\hspace{-1.5mm}[1,1,-1]_0^+ +
_{-2}^{\hspace{1.4mm}+}\hspace{-1mm}[2,0,0]_0^+\\
&= R_{-2,2}(_{-1}^{\ +}\hspace{-0.5mm}[1,1,-1]_{-2}^+)
\hspace{1mm}\#\hspace{1mm} _{\hspace{0.4mm}0}^+\hspace{-0.5mm}[2,0,0]_0^+\\
&= _{-3}^{\hspace{1.4mm}+}\hspace{-1.5mm}[-1,1,1]_0^+
\hspace{1mm}\#\hspace{1mm} _{\hspace{0.4mm}0}^+\hspace{-0.5mm}[2,0,0]_0^+\\
&= _{-3}^{\hspace{1.4mm}+}\hspace{-1.5mm}[1,1,1]_0^+.
\end{align*}
However, we have
$$
R_{2,-2}(_{-3}^{\hspace{1.4mm}+}\hspace{-0.2mm}[1,1,1]_0^+)=_{-1}^{\
+}\hspace{-1.5mm}[P^+_2(1,1,1)]_{-2}^+=_{-1}^{\
+}\hspace{-1.5mm}[1,1,1]_{-2}^+,
$$ satisfying Theorem \ref{theoInt}. Interestingly, $B_2+B_1$ does
not work because while $B_2$ and $B_1$ have the same end-node state,
the common twist $-1$ is an odd number, which violates the condition
of interaction.

Equipped with this algebra, we shall prove our primary result:
\begin{corollary}
Given two actively interacting braids $B=\left. _{T_l}^S\hspace{-0.5
mm}[ T_a,T_b,T_c] _{T_r}^S\right.$ and $B'=\left. _{\ T_l'}^{\pm
S}\hspace{-0.5 mm}[ T'_a,T'_b,T'_c] _{T'_r}^{\pm S}\right.$
satisfying the interaction conditions, the resulting braid $B^f = B
+ B'$ conserves the quantities $T_l+T_r+T'_l+T'_r = T^f_l+T^f_r$,
$\sum\limits_{i=a}^{c}(T_i+T'_i) = \sum\limits_{i=a}^{c}T^f_i$ and
$S^2$.
\end{corollary}

\begin{proof}
We immediately use the result from the proof of the previous theorem:
\begin{equation}
B^f = \left. _{T_l+T_r+T'_l}^{\hspace{9mm} \pm S}\hspace{-0.5 mm}[
(P^S_{T_r+T_l'}(T_a,T_b,T_c))+(T'_a,T'_b,T'_c)] _{T'_r}^{\pm
S}\right.
\end{equation}
Because this result is independent of the form of the braids before
the interaction according to the previous theorems, we need only
examine the result of the interaction. $S^2$ is conserved as our
final result shares the same value for $S^2$ of each of our initial
states, and $S^2$ is invariant under the equivalence moves. We know
that $T^f_l+T^f_r$ is conserved under the equivalence moves, and
from our result we see that this conserved quantity is equal to
$T_l+T_r+T'_l+T'_r$. Finally, $\sum\limits_{i=a}^{c}T^f_i$ is the
same regardless of the form of $P^S_{T_r+T_l'}$, this immediately
has the form of $\sum\limits_{i=a}^{c}(T_i+T'_i)$.
\end{proof}

The three conserved quantities of this theorem have a clear meaning:
\begin{enumerate}
\item $T_l+T_r+T'_l+T'_r = T^f_l+T^f_r$, the total external twists before and after an interaction
are the same;
\item $\sum\limits_{i=a}^{c}(T_i+T'_i)$, the total internal twists
remains the same under interaction;
\item $S^2$, the interacting character of the braids is preserved.
\end{enumerate}

\section{Conclusions}
Conservation laws are a valuable tool in gaining understanding of
the underlying structure of a theory. Elucidating the actual content
of a theory and revealing information that could otherwise remain
inaccessible. Equipped with invariants and conserved quantities, we
are able to work to determine how the content of the theory relates
to particle physics.

We have developed an algebraic notation for our braids, found a set
of equivalence relations relating them, and developed conserved
quantities associated with these relations. More importantly, we
have found the relationship between these algebraic forms and the
interaction of actively-interacting braids. From this we have found
quantities conserved under interaction. These are dynamically
conserved quantities, which are also conserved under the equivalence
moves.

This allows us to work towards a better understanding of and a
classification of the complete set of these braids. That we are able
to show that actively-interacting braids interact with each other to
produce actively-interacting braids allows us to generate an
infinite set of such braids.

The next step in this work is to determine which of the conserved
quantities found in this paper may be mapped to quantum numbers
characterizing fundamental particles and to use the results we have
found for interactions of braids to fully classify the set of
braids.

\section*{Acknowledgements}

The authors are indebted to their Advisor, Lee Smolin, for his
encouragement and critical comments. We thank Sudance
Bilson-Thompson for helpful discussions. Research at Perimeter
Institute for Theoretical Physics is supported in part by the
Government of Canada through NSERC and by the Province of Ontario
through MRI.

\end{document}